\documentclass{PoS}

\usepackage{graphicx}

\usepackage{amssymb}
\usepackage{epsfig}
\usepackage{amssymb}
\usepackage{epsf}
\usepackage{epsfig}
\usepackage{color}
\usepackage{ifpdf}
\usepackage{longtable}
\usepackage{cite}

\newcommand{\nn}{\nonumber}
\newcommand{\ba}{\begin{eqnarray}}
\newcommand{\ea}{\end{eqnarray}}
\newcommand{\be}{\begin{equation}}
\newcommand{\ee}{\end{equation}}
\newcommand{\bd}{\begin{displaymath}}
\newcommand{\ed}{\end{displaymath}}

\newcommand{\old}[1]{}

\newcommand{\myangle}{0}
\newcommand{\mywidth}{200pt}

\title{The neutral pion decay and the chiral anomaly on the lattice}

\ShortTitle{The neutral pion decay and the chiral anomaly on the lattice}

\author{
\vspace{-0.05\textheight}
\begin{flushleft}
\hspace{0.75\textwidth}KEK-CP-281\\
\hspace{0.75\textwidth}UTHEP-651
\end{flushleft}
JLQCD Collaboration:
\speaker{X. Feng}$^a$\thanks{E-mail: xufeng@post.kek.jp},
 S. Aoki$^{b,c}$,
 H. Fukaya$^d$,
 S. Hashimoto$^{a,e}$,
 T. Kaneko$^{a,e}$,
 J. Noaki$^a$ and E. Shintani$^f$
 \\ \\
 \\
\llap{$^a$}High Energy Accelerator Research Organization (KEK), Ibaraki 305-0801
, Japan\\
\llap{$^b$}Graduate School of Pure and Applied Sciences, University of Tsukuba, 
Ibaraki 305-8571, Japan\\
\llap{$^c$}Center for Computational Sciences, University of Tsukuba, Ibaraki 305-8577, Japan\\
\llap{$^d$}Department of Physics, Osaka University, Toyonaka, Osaka 560-0043, Japan\\
\llap{$^e$}School of High Energy Accelerator Science, The Graduate University for Advanced Studies (Sokendai), Ibaraki 305-0801, Japan\\
\llap{$^f$}RIKEN-BNL Research Center, Brookhaven National Laboratory, Upton, NY 11973, USA}

\abstract{
We perform a lattice QCD calculation of the $\pi^0\to\gamma\gamma$
transition form factor and the associated decay width.
We use a Euclidean time integral of
the relevant three-point function to compute the decay amplitude 
for two-photon final state, which is a non-QCD state.
We use the all-to-all quark
propagator technique to carry out this integral as well as to include 
the disconnected quark diagram contributions. 
We execute the calculation using the overlap fermion formulation, which
ensures the exact chiral
symmetry on the lattice and produces the chiral anomaly through
the Jacobian of the chiral transformation.
We examine various sources of systematic
effects except for possible discretization effect.
Our final results for the form factor and the decay width
reproduce the ABJ anomaly in the chiral limit and
agree with the experimental measurements at the physical pion mass with a precision
of a few percent. 
}

\FullConference{The 30 International Symposium on Lattice Field Theory - Lattice 2012,\\
                June 24-29, 2012\\
                Cairns, Australia}

\begin{document}

\section{Introduction}
The pions are supposed to be the Nambu-Goldstone bosons 
associated with the spontaneous chiral symmetry breaking in QCD.
The three pions ($\pi^+$, $\pi^0$, $\pi^-$) form an isospin triplet
of flavor SU(2) symmetry. Among the three pions, 
$\pi^0$ is most unstable one, 
with a lifetime $\sim10^{-9}$ times shorter than that of the other two.
Experiments show that
the neutral pion decay is mainly an electromagnetic process, with 
the bulk of the decay rate going to two photons.

At the leading order of QED
the $\pi^0\rightarrow\gamma\gamma$ decay width
can be expressed as
\ba
\label{eq:width}
\Gamma_{\pi^0\gamma\gamma}=\frac{\pi\alpha_e^2m_\pi^3}{4}{\mathcal F}^2_{\pi^0\gamma\gamma}(m_\pi^2,0,0)\;,
\ea
where $\alpha_e$ is the fine structure constant, $m_\pi$ is the neutral pion mass and ${\mathcal F}_{\pi^0\gamma\gamma}(m_\pi^2,p_1^2,p_2^2)$ is the $\pi^0\rightarrow\gamma\gamma$ transition form factor with $p_{1,2}$ the photon momenta.
In 1967 Sutherland and Veltman
argued that in the chiral limit
${\mathcal F}_{\pi^0\gamma\gamma}(0,0,0)$ vanishes and the pion can not
decay into photons~\cite{Sutherland:1967vf,Veltman}. At the physical pion mass
the predicted decay width is 
$\sim$1000 times smaller than the experimental 
measurements. 
Later it was realized that the PCAC relation used in Sutherland and Veltman's analysis
is only valid at the classical level. In quantum field theory, the conservation
of the axial current
is violated by quantum fluctuations. At the one-loop level, a fermionic 
triangle diagram contributes an extra anomaly term ({\em ABJ anomaly}) 
to the PCAC relation
and makes ${\mathcal F}_{\pi^0\gamma\gamma}(m_\pi^2,0,0)$ non-zero in the 
chiral limit~\cite{Adler:1969gk,Bell:1969ts}
\ba
\label{eq:abj}
{\mathcal F}_{\pi^0\gamma\gamma}^{\rm ABJ}\equiv{\mathcal F}_{\pi^0\gamma\gamma}(0,0,0)=\frac{N_c}{12\pi^2F_0}\;.
\ea
In Eq.~(\ref{eq:abj}) $N_c$ is the number of QCD colors and 
$F_0$ is the pion decay constant $F_\pi$ in the chiral limit.
It is proved in perturbation theory
that higher-loop diagrams do not contribute to ${\mathcal F}_{\pi^0\gamma\gamma}(0,0,0)$~\cite{Adler:1969er}. As a consequence, the ABJ anomaly
gives a rather precise predication for the $\pi^0\rightarrow\gamma\gamma$ decay rate.


In the chiral and on-shell photon limit, the pion decay amplitude is constrained by the ABJ anomaly. Away from these limits some corrections from QCD are expected.
The recent PrimEx experiment at JLab has measured
the neutral pion decay width 
to an accuracy of 2.8\%~\cite{Larin:2010kq}. The next stage of this experiment 
is to achieve a precision of 1.4\%.
At this level of accuracy the correction due to finite quark mass
becomes relevant.
In this paper we report a model-independent calculation 
of the $\pi^0\rightarrow\gamma\gamma$
form factor and decay width using lattice QCD. 
The first motivation of this work is to determine 
the finite quark mass correction from first principles.
Our second motivation comes from the hadronic
light-by-light (HLbL) scattering, which is responsible for the second largest
theoretical error in the determination of the muon g-2. 
While the direct QCD calculation
of HLbL is very demanding as it involves a 
four-vector-current correlation function,
the $\pi^0\rightarrow\gamma\gamma$ form factor
can be used to estimate the dominant pion-exchange contribution 
to the HLbL. Thus our calculation serves
as an intermediate step towards the precise determination of HLbL.

\section{Chiral anomaly on the lattice} 
The chiral anomaly is of central importance for the neutral pion decay. 
When we perform a lattice calculation, a natural question is
how the chiral anomaly is achieved on the lattice.
The answer depends on the formulation of the fermion action.
In the case of Wilson fermion, the chiral symmetry is explicitly violated and 
the anomaly is recovered by taking the continuum limit of the chiral symmetry 
breaking term~\cite{Karsten:1980wd}.
For the Ginsparg-Wilson fermions, the chiral symmetry is
preserved in a modified form~\cite{Luscher:1998pqa}. 
The chiral anomaly is introduced by the Jacobian of
the chiral transformation in this case. When the background gauge field is 
sufficiently smooth, the Jacobian yields the correct chiral anomaly 
up to discretization effects~\cite{Fujikawa:2000qw}. 

In our calculation we use the overlap fermion formulation, which is
a realization of the Ginsparg-Wilson fermion on the lattice. At practically
used lattice spacings ($\sim$0.1 fm) the gauge field is far from
smooth and the chiral anomaly may not be guaranteed. Therefore it is
important to check whether the chiral anomaly
is correctly reproduced in our calculation.

\section{Treatment of non-QCD state}
Lattice QCD provides a powerful tool to calculate the matrix elements with
hadronic initial/final state. By studying the Euclidean time dependence 
of the correlation function, we are able to pick up the
hadronic state of interest. 
However, this method does not work for the matrix elements with non-QCD state.
Take the photon state as an example. By using the interpolating operator with quantum number $J^{PC}=1^{--}$, we expect to extract vector-meson state rather than the photon state from the correlation function.
To address this problem, Ji and Jung proposed an
analytic continuation method, which treats the photon as 
a superposition of a complete set of hadron states with appropriate
quantum numbers~\cite{Ji:2001wha}.
In our calculation the key observable is
the matrix element $\langle \gamma(p_1,\lambda_1)\gamma(p_2,\lambda_2) | \pi^0(q)\rangle
$ with a two-photon
final state, for which we apply this technique.

We follow the procedure given in Refs.~\cite{Dudek:2006ut,Cohen:2008ue}.
Using the LSZ reduction formula, we express the matrix element in terms of
the time-ordered correlation function
\ba
\label{eq:LSZ}
\langle \gamma(p_1,\lambda_1)\gamma(p_2,\lambda_2) | \pi^0(q)\rangle
=&&- \lim_{p'_{1,2} \to p_{1,2}} \epsilon^*_\mu(p_1,
 \lambda_1) \epsilon^*_\nu(p_2, \lambda_2)\nn\\
&&\hspace{-2cm}\times p_1'^2 p_2'^2 \int d^4x\; d^4y \; e^{i p'_1x +
 i p'_2y} \langle 0 | T\big\{ A^\mu(x) A^\nu(y) \big\} | \pi^0(q) \rangle\;.
\ea
At the leading order of perturbative QED expansion, the photon field 
in the interaction term $H_{int}=e\int d^4x\;A^\mu(x)j_\mu(x)$ 
can be contracted with these photon fields
existing in Eq.~(\ref{eq:LSZ}). After the Wick contraction we have
\ba
\label{eq:pQED}
\langle \gamma(p_1,\lambda_1)\gamma(p_2,\lambda_2) | \pi^0(q)\rangle
=&&-e^2 \lim_{p'_{1,2} \to p_{1,2}} \epsilon^*_\mu(p_1,
 \lambda_1) \epsilon^*_\nu(p_2, \lambda_2)\nn\\
&&\hspace{-4cm}\times p_1'^2 p_2'^2 \int d^4x\; d^4y \;d^4z\;d^4w\; 
e^{i p'_1x +
 i p'_2y} D^{\mu\rho}(x,z) D^{\nu\sigma}(y,w)  \langle \Omega | T\big\{ j_\rho(z) j_\sigma(w) \big\} | \pi^0(q) \rangle\;.
\ea
In Eq.~(\ref{eq:pQED}) $j_{\rho}$ and $j_{\sigma}$ are the hadronic
components of the electromagnetic vector current.
The photon propagator $D^{\mu\rho}(x,z)=-ig^{\mu\rho}\int\frac{d^4k}{(2\pi)^4}\frac{e^{-ik(x-z)}}{k^2+i\epsilon}$ cancels the inverse propagators outside the
integral. We then have 
\ba
\label{eq:def_Minkowski}
&&\langle \gamma(p_1,\lambda_1)\gamma(p_2,\lambda_2) | \pi^0(q)\rangle
=-ie^2\epsilon^*_\mu(p_1,\lambda_1) \epsilon^*_\nu(p_2, \lambda_2)M_{\mu\nu}(p_1,p_2)\;,\nn\\
&&M_{\mu\nu}(p_1,p_2)=i\int d^4x\;e^{ip_1x}\langle\Omega|T\{j_\mu(x)j_\nu(0)\}|\pi^0(q)\rangle\;.
\ea
where $M_{\mu\nu}(p_1,p_2)$ is a hadronic matrix element.
The form factor ${\mathcal F}_{\pi^0\gamma\gamma}(m_\pi^2,p_1^2,p_2^2)$ is defined in terms of $M_{\mu\nu}(p_1,p_2)$ as
\ba
{\mathcal F}_{\pi^0\gamma\gamma}(m_\pi^2,p_1^2,p_2^2)
=M_{\mu\nu}(p_1,p_2)/\varepsilon_{\mu\nu\alpha\beta}p_1^\alpha p_2^\beta\;,
\ea
where the factor $\varepsilon_{\mu\nu\alpha\beta}p_1^\alpha p_2^\beta$ is induced by the negative parity of $\pi^0$.

By an analytic continuation of
(\ref{eq:def_Minkowski}) from the Minkowski to Euclidean space-time,
we write
\ba
\label{eq:correlator}
&&M_{\mu\nu}(p_1,p_2)=\lim_{t_{1,2}-t_\pi\rightarrow\infty}\frac{1}{\frac{\phi_{\pi,\vec{q}}}{2E_{\pi,\vec{q}}}e^{-E_{\pi,\vec{q}}(t_2-t_\pi)}}
\int dt_1\;e^{\omega (t_1-t_2)}C_{\mu\nu}(t_1,t_2,t_\pi)\;,\\
&&
\label{eq:correlator1}
C_{\mu\nu}(t_1,t_2,t_\pi)\equiv\int d^3\vec{x}\;e^{-i\vec{p}_1\cdot\vec{x}}
\int d^3\vec{z}\;e^{i\vec{q}\cdot\vec{z}}
\langle\Omega|T\{j_\mu(\vec{x},t_1)j_\nu(\vec{0},t_2)\pi^0(\vec{z},t_\pi)\}|\Omega\rangle\;,
\ea
where $C_{\mu\nu}(t_1,t_2,t_\pi)$
is a correlation function defined in Euclidean space-time and thus can be
 calculated using lattice QCD. The operator 
$\int {d^3\vec{z}}\;e^{i\vec{q}\cdot\vec{z}}\pi^0(\vec{z},t_\pi)$
produces a neutral pion with a spatial momentum $\vec{q}$.
Its amplitude and energy in the ground state are denoted by $\phi_{\pi,\vec{q}}$ and
$E_{\pi,\vec{q}}$, respectively. The four-momentum of the first photon $p_1=(\omega,\vec{p}_1)$
is chosen as input, while the momentum of the second photon is given
 as $p_2=(E_{\pi,\vec{q}}-\omega,\vec{q}-\vec{p}_1)$ by momentum conservation.
When the squared momentum $p_1^2$ or $p_2^2$ exceeds the hadron production threshold,
the photon state mixes with these hadron states and the analytic
continuation fails.
To avoid this, we restrict the photon momentum in the region
$p_{1,2}^2<M_V^2$, where $M_V$ is the invariant mass of the lowest energy state in the vector channel. Although $e^{\omega(t_1-t_2)}$ becomes infinitely large when $t_1-t_2\rightarrow\infty$, a suppression factor $e^{-E_V(t_1-t_2)}$ from $C_{\mu\nu}(t_1,t_2,t_\pi)$
makes the integratal~(\ref{eq:correlator}) convergent.

\section{Lattice setup}
In this calculation we use 2+1-flavor
overlap fermion configurations generated by
the JLQCD and TWQCD Collaborations~\cite{Matsufuru:2008zz,Noaki:2010zz}.
Using the overlap fermions ensures the exact chiral symmetry at even
finite lattice spacings. We use a sequence of ensembles
with a lattice spacing of $a=0.11$ fm.
The pion mass ranges from 290 to 540 MeV with degenerate up and down quarks.
The strange quark mass
is fixed to be very close to its estimated physical value.
The lattice size is $L^3\times T/a^4=16^3\times 48$.
At two smallest pion masses we also use 
a larger lattice size $L/a=24$ to check the finite-size (FS) effects. 
The gauge ensembles are generated by fixing the (global) topological charge
$Q$, which results in a finite volume effect of $O(1/L^3T)$~\cite{Aoki:2007ka}. 
We check the significance of this effect by comparing the results with 
two different values $Q=0,1$.

We use the all-to-all propagator to construct the correlation function.
Since the propagator contains the information
from any source site to any sink site, we are allowed to
calculate $C_{\mu\nu}(t_1,t_2,t_\pi)$ at
any time slices of $t_1$, $t_2$ and $t_\pi$.
Besides, we are able to compute the disconnected diagrams
without extra computer resources.
The electromagnetic current $j_\mu$ is
implemented on the lattice as
a local operator with a renormalization factor calculated nonperturbatively in~\cite{Noaki:2009xi}.

\section{Analysis}
\begin{figure}
\begin{minipage}[t]{210pt}
\includegraphics[width=\mywidth,angle=\myangle]{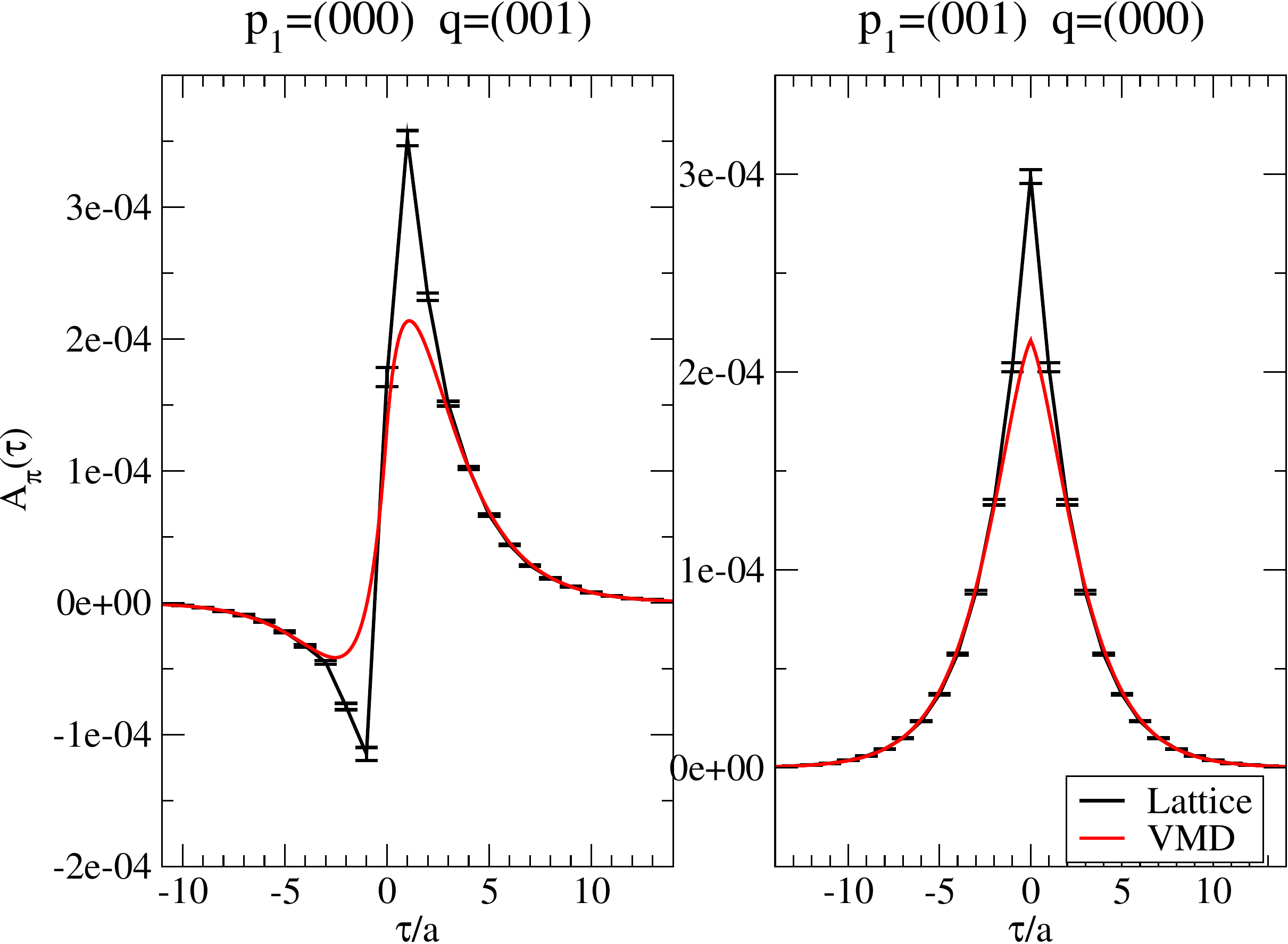}
\caption{
The amplitude $A_{\pi}(\tau)$ as a function of $\tau$ for momentum setups 1 (left) and 2 (right).
The black (red) curves indicate the lattice (VMD) amplitudes.
}
\label{fig:distribution}
\end{minipage}
\hspace{4pt}
\begin{minipage}[t]{210pt}
\includegraphics[width=\mywidth,angle=\myangle]{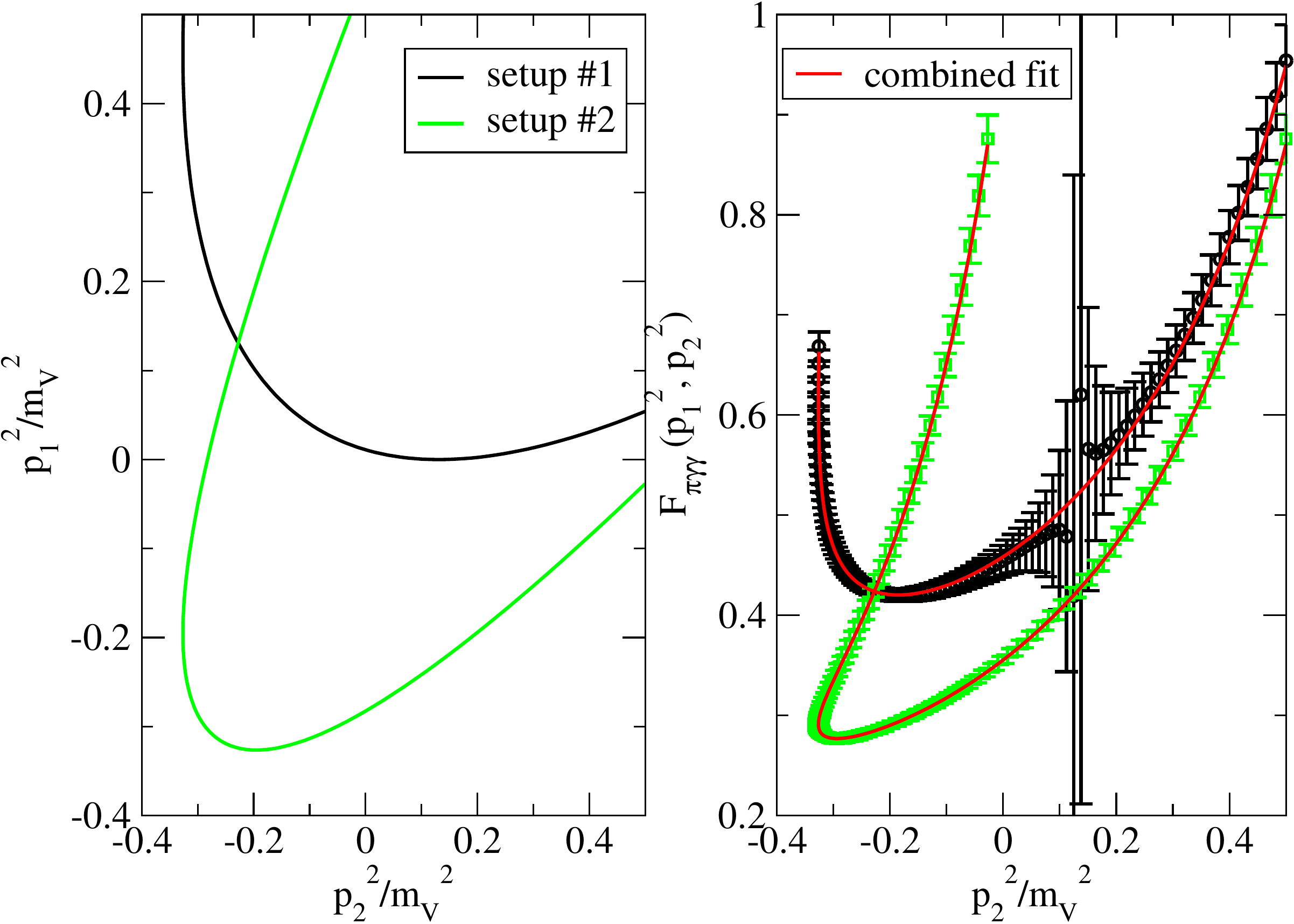}
\caption{
Left: Contour of $(p_1^2,p_2^2)$ rescaled by $1/M_V^2$ for our momentum setups.
Right: $F_{\pi^0\gamma\gamma}(m_\pi^2,p_1^2,p_2^2)$
as a function of $p_2^2/M_V^2$. 
}
\label{fig:matrix_element}
\end{minipage}
\end{figure}

From the large $t_{1,2}-t_\pi$ behavior of $C_{\mu\nu}(t_1,t_2,t_\pi)$,  
it is possible to extract the $\pi^0$-ground state.
We define the amplitude $A_\pi$ as
\ba
A_\pi(\tau)\equiv\lim_{t-t_\pi\rightarrow\infty}C_{\mu\nu}(t_1,t_2,t_\pi)/e^{-E_{\pi,\vec{q}}(t-t_\pi)}\;,
\ea
with $\tau=t_1-t_2$ and $t=\textmd{min}\{t_1,t_2\}$, and obtain $M_{\mu\nu}(p_1,p_2)$ by performing an integral
\ba
\label{eq:R_expression}
&&\hspace{-0.6cm}\frac{2E_{\pi,\vec{q}}}{\phi_\pi}\left(\int_0^{\infty} d\tau\;e^{\omega \tau}A_{\pi}(\tau)
+\int_{-\infty}^0 d\tau\;e^{(\omega-E_{\pi,\vec{q}}) \tau}A_{\pi}(\tau)\right)\;.
\ea
The spatial momenta are assigned as
$\vec{p}_1=\frac{2\pi}{L}(0,0,0)$, $\vec{q}=\frac{2\pi}{L}(0,0,1)$ (setup 1)
and
$\vec{p}_1=\frac{2\pi}{L}(0,0,1)$, $\vec{q}=\frac{2\pi}{L}(0,0,0)$ (setup 2).
The resulting amplitudes $A_{\pi}(\tau)$ for these setups
are shown by the black curves in Fig.~\ref{fig:distribution}.
The amplitude $A_{\pi}^{\rm VMD}(\tau)$, constructed from the vector-meson-dominance (VMD) form factor
${\mathcal F}_{\pi^0\gamma\gamma}^{\rm VMD}(m_\pi^2,p_1^2,p_2^2)=c_VG_V(p_1^2)G_V(p_2^2)$, with $G_V(p^2)=M_V^2/(M_V^2-p^2)$ the vector meson propagator and $c_V$ a constant is also plotted by red curves.
We find that $A_{\pi}^{\rm VMD}(\tau)$ give a good approximation 
to the lattice data
at large $|\tau|$ but fails to match them at small $|\tau|$. 
This is because no information of the vector-meson excited states
are contained in ${\mathcal F}_{\pi^0\gamma\gamma}^{\rm VMD}(m_\pi^2,p_1^2,p_2^2)$.
Although the lattice data are truncated due to 
the finite time extent $T$, we are 
still able to perform the integral~(\ref{eq:R_expression}) 
from $-\infty$ to $+\infty$ since the 
$A_{\pi}(\tau)$ at the large $|\tau|$ is dominated by the lowest vector meson.

\begin{figure}[tb]
\begin{center}$
\begin{array}{c}
\includegraphics[width=\mywidth,angle=\myangle]{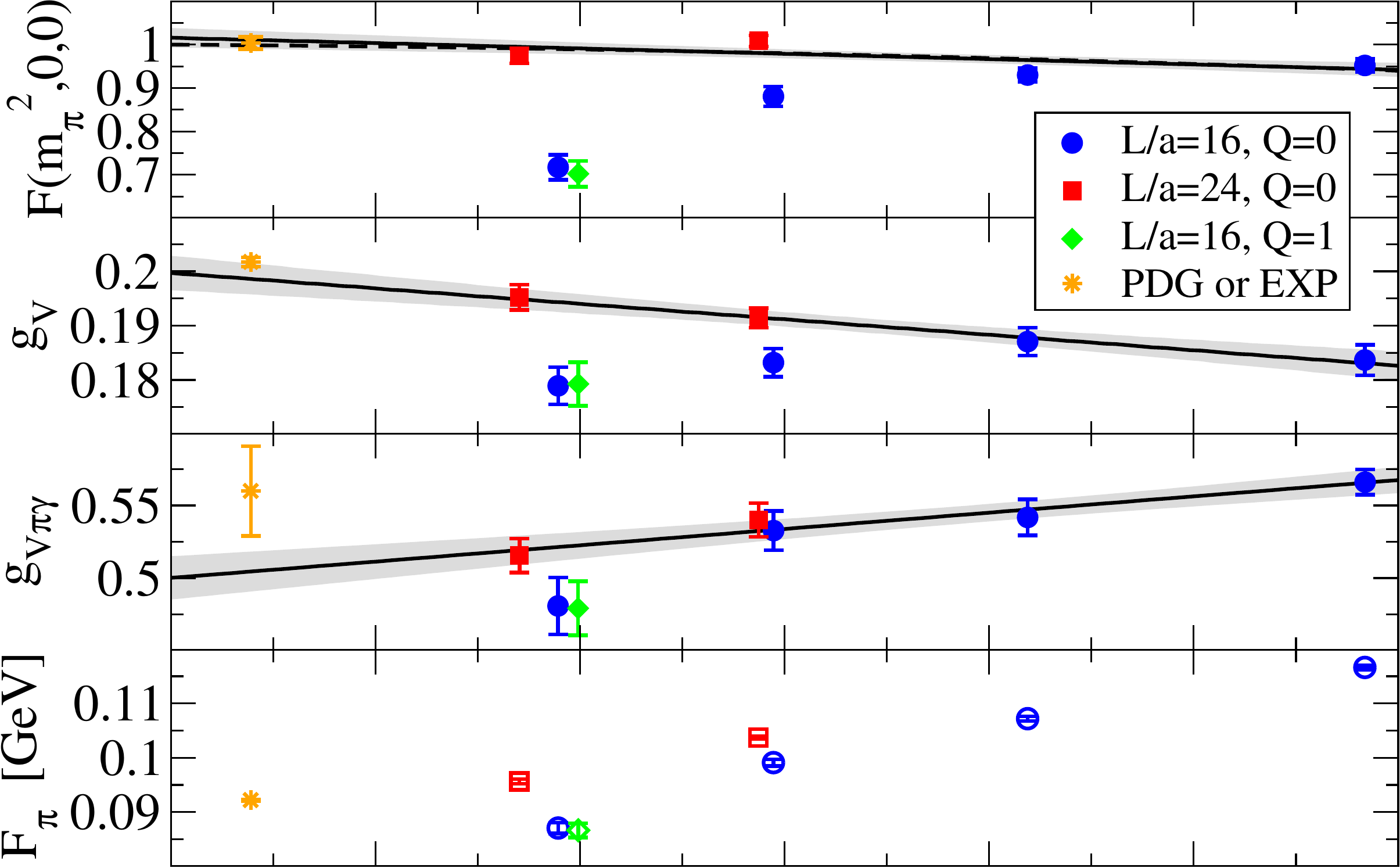} \\
\includegraphics[width=\mywidth,angle=\myangle]{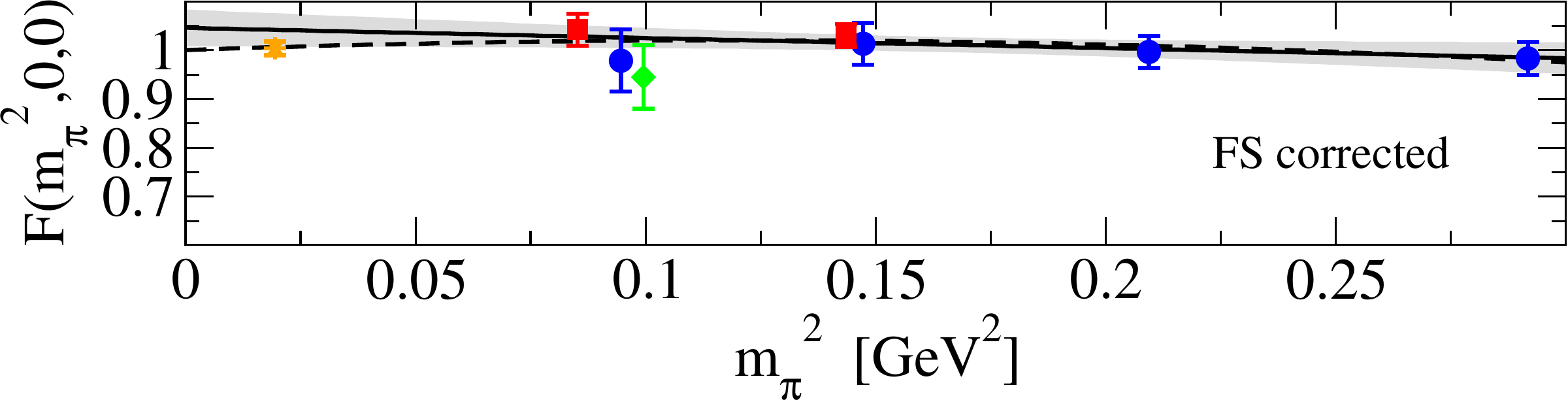}
\end{array}$
\end{center}
\caption{
$F(m_\pi^2,0,0)$, $g_V$, $g_{V\pi\gamma}$, $F_\pi$ and FS corrected $F(m_\pi^2,0,0)$ as a function of $m_\pi^2$ from top to bottom panels. In each panel, data with $(L/a,Q)=(16,0)$, $(24,0)$ and $(16,1)$ are plotted by the blue, red and green symbols, respectively. The yellow symbols indicate the Particle Data Group (PDG)~\cite{Nakamura:2010zzi} or PrimEx~\cite{Larin:2010kq} experimental values for reference.
The solid (dashed) curves show the result of the fit to the linear (quadratic) function.
The dataset used in the fit is explained in the text.
}
\label{fig:ff_wo_FS}
\end{figure}

When performing the integral~(\ref{eq:R_expression}) we can tune the value of the photon energy
$\omega$ continuously.
As shown in the left panel of Fig.~\ref{fig:matrix_element}, a pair $(p_1^2,p_2^2)=(\omega^2-\vec{p}_1^2,
(E_{\pi,\vec{q}}-\omega)^2-(\vec{q}-\vec{p}_1)^2)$ forms a continuous contour on
the $(p_1^2,p_2^2)$ plane for $p_{1,2}^2<M_V^2/2$.
Evaluating ${\mathcal F}_{\pi^0\gamma\gamma}(m_\pi^2,p_1^2,p_2^2)$ along this contour,
we obtain the data plotted in the right panel of Fig.~\ref{fig:matrix_element}.
We use the fit ansatz ${\mathcal F}_{\pi^0\gamma\gamma}(m_\pi^2,p_1^2,p_2^2)=$
\ba
\label{eq:expansion}
c_VG_V(p_1^2)G_V(p_2^2)
+\sum_{m} c_m
\left((p_2^2)^mG_V(p_1^2)+(p_1^2)^mG_{V}(p_2^2)\right)+\sum_{m,n}c_{m,n}(p_1^2)^m(p_2^2)^n\;,
\ea
which includes the contribution from the lowest vector meson 
through $G_V(p_{1,2}^2)$ and
the contribution from excited states as a polynomial of $p_{1,2}^2$.
We perform the combined fit of the lattice data to Eq.~(\ref{eq:expansion}) with four free parameters: $c_V$, $c_0$,
$c_{0,0}$ and $c_{0,1}=c_{1,0}$. The contributions from the higher-order terms
are not significant when compared to the statistical error and thus can be neglected.
The fitting curves are shown in the right panel
of Fig.~\ref{fig:matrix_element}.
We find that the single formula~(\ref{eq:expansion}) describes the data with different 
momentum setups.
Using the resulting fit parameters and extrapolating
${\mathcal F}_{\pi^0\gamma\gamma}(m_\pi^2,p_1^2,p_2^2)$ to the soft photon limit,
 we obtain 
the normalized form factors $F(m_\pi^2,0,0)=(4\pi^2F_\pi){\mathcal F}_{\pi^0\gamma\gamma}(m_\pi^2,0,0)$, which are plotted in the uppermost panel of 
Fig.~\ref{fig:ff_wo_FS}.

Using the results of $F(m_\pi^2,0,0)$ we analyze the various systematic effects including FS effects, fixed-topology effects, disconnected-diagram effects, higher-order effects in 
chiral extrapolation and possible discretization effects
in the evaluation of Eq.~(\ref{eq:correlator}). 
For more details of the analysis, we refer the readers to our recent 
publication~\cite{Feng:2012ck}.

Among the various systematic effects, the largest one is the conventional FS effect. As shown in Fig.~\ref{fig:ff_wo_FS}, 
 at $m_\pi\approx 290$ MeV we find that $F(m_\pi^2,0,0)$
 calculated at $L/a = 16$ lattice is 27\% less than the one at $L/a = 24$.
To qualitatively understand this FS effect, we
insert the ground state into $\langle j_\mu j_\nu \pi^0\rangle$ and
approximate this three-point correlation function with three hadronic matrix elements:
$\langle j_\mu j_\nu \pi^0\rangle \rightarrow \langle \Omega|j_\mu|V,\varepsilon\rangle\langle V,\varepsilon |j_\nu |\pi^0\rangle\langle\pi^0|\pi^0|\Omega\rangle$.
These matrix elements are related to the electromagnetic coupling $g_V$,
the $V\pi\gamma$ coupling $g_{V\pi\gamma}$
and the pion decay constant $F_\pi$.
In our calculation we do not observe significant FS effect in $M_V$ but
find 8\%, 7\% and 9\% shifts in $g_V$, $g_{V\pi\gamma}$ and $F_\pi$, respectively, from $L/a=16$ to 24, as shown in Fig.~\ref{fig:ff_wo_FS}.
These FS effects may accumulate in the three-point function.
We estimate the FS corrections to
$g_V$, $
g_{V\pi\gamma}$ and $F_\pi$ by adding a correction term, $e^{-m_\pi L}$, into the
linear fit form in the chiral extrapolation of each quantity or using NNLO chiral
perturbation theory. We then combine the FS corrections to each quantity as an estimation of the total FS correction to
$F(m_\pi^2,0,0)$.
As shown in the lowest panel of Fig.~\ref{fig:ff_wo_FS} the
FS corrected $F(m_\pi^2,0,0)$ at $L/a=16$ agrees with
those at $L/a=24$.

We use two methods of treating the FS effects: 1. We use the uncorrected $F(m_\pi^2,0,0)$ with
$m_\pi L>4$ to perform the chiral extrapolation. Namely, we exclude the $L/a=16$ data points at two lowest pion masses. A linear function in $m_\pi^2$
is used as an fit ansatz.  2. We use the FS corrected data of all ensembles to perform a linear extrapolation. The difference between the results from the two methods is
considered as a systematic error.

We quote our results for $F(0,0,0)$ and $\Gamma_{\pi^0\gamma\gamma}$ in the isospin symmetric limit as
$F (0,0,0)= 1.009(22)(29)$ and
$\Gamma_{\pi^0\gamma\gamma}=7.83(31)(49)$ eV,
where the first error is statistical and the second one is systematic.
Our results reproduce the predication of the ABJ anomaly $F(0,0,0)=1$ and
agree with the PrimEx measurement $\Gamma_{\pi^0\gamma\gamma}=7.82(22)$ eV~\cite{Larin:2010kq}.
For future improvements, 
isospin breaking effects due to the light quark mass difference need to be included.

\vspace{3mm}
Numerical simulations are performed on the Hitachi SR16000 at Yukawa Institute of Theoretical Physics and
at High Energy Accelerator Research Organization
under a support of its Large Scale Simulation Program
(No. 11-05). This work is supported in part by the Grant-in-Aid of
the Japanese Ministry of Education (No.21674002, 21684013),
the Grant-in-Aid for Scientific Research on Innovative Areas
(No. 2004: 20105001, 20105002, 20105003, 20105005, 23105710), and
SPIRE (Strategic Program for Innovative Research).

\bibliography{lat12_xufeng}
\bibliographystyle{h-physrev}

\end{document}